\documentclass{sig-alternate-10pt}

\pdfoutput=1
\usepackage{times}
\usepackage{graphicx}
\usepackage{multirow}
\usepackage{subfigure}
\usepackage{url}
\usepackage{amssymb,amsmath}
\usepackage{epstopdf}
\usepackage{algorithm}
\usepackage{algpseudocode}

\title{Multipath forwarding strategies in Information Centric Networks with AIMD congestion control}

\numberofauthors{1}
\author{
\alignauthor
Andrea Detti,  Claudio Pisa, Nicola Blefari-Melazzi\\
\affaddr{CNIT University of Rome Tor Vergata}
\email{name.surname@uniroma2.it}
}

\begin{document}
\maketitle

\begin{abstract}
A content can be replicated in more than one node, in Information Centric Networks (ICNs). Thus, more than one path can be followed to reach the same content, and it is necessary to decide the interface(s) to be selected in every network node to forward content requests towards such multiple content containers. A multipath forwarding strategy defines how to perform this choice. 
In this paper we propose a general analytical model to evaluate the effect of multipath forwarding strategies on the performance of an ICN content delivery, whose congestion control follows a receiver driven, loss-based AIMD scheme. We use the model to understand the behavior of ICN multipath forwarding strategies proposed in the literature so far, and to devise and evaluate a novel strategy. The considered multipath forwarding strategies are also evaluated in a realistic network setting, by using the PlanetLab testbed.
\end{abstract}




\section{Introduction}
The Internet today is more and more used as a container of information in which users can put content or from which they can get content.
Correspondingly, networks must adopt efficient solutions to distribute contents, rather than to create host-to-host bit pipes.
Efficient content distribution and dissemination systems exploit network strategies that try to jointly optimize communication, storage and computation resources. Content replication, caching, content routing, content adaptation are typical functionality of e.g. P2P applications, Content Delivery Networks and Information Centric Networks (ICNs).

ICN is an emerging network paradigm that rethinks network services by putting information delivery at the center of the network layer design. Whereas the current Internet model aims to create network pipes between hosts identified by addresses, ICN delivers to the users information (or contents) identified by names. A user expresses an interest for a content and the ICN functionality takes care of routing the content request towards the best source (be it the original one, a replica server, or an in-network cache) and of sending back to the user the requested data. Content Centric Network (CCN) \cite{jacobson2009networking} is probably the best-known among the proposed ICN architectures \cite{polyzos}. A CCN includes routing-by-name,  multicast delivery, receiver-driven congestion control and in-network caching functionality.

In-network caching and/or possible content replication results in the same content being available in multiple locations of an ICN. Thus, multipath solutions able to exploit the available paths are very useful to speed up delivery and improve resilience \cite{carofigliomultipath} in ICN.

A full multipath solution, either based on TCP/IP or ICN, usually includes: i) \emph{path discovery}, ii) \emph{congestion control}, and iii) \emph{multipath forwarding} mechanisms. The path discovery makes involved nodes aware of the existence of multiple paths towards a given content. The congestion control regulates the data flow on the selected multiple paths. The multipath forwarding schedules traffic among available paths according to a given strategy; it can operate either on a per-packet basis or on a per-flow basis.

A per-packet forwarding strategy selects a forwarding path packet-by-packet. Thus, the packets of a given flow concurrently exploits all available paths and the achievable overall transfer rate is bounded by the sum of the rates available on all paths.
A per-flow strategy selects the forwarding path flow-by-flow, where a flow is the sequence of packets related to an end-to-end activity, e.g. a TCP connection or a ICN content delivery session. All the packets of a flow follow the same path, while packets of different flows may use different paths. Therefore, the transfer rate of a flow is bounded by the rate available on the selected path.

In this paper we present \textbf{a general analytical model} to evaluate the effect of per-packet multipath forwarding strategies on the performance of an ICN content delivery, whose congestion control is regulated by a receiver-driven, loss-based Additive Increase Multiplicative Decrease (AIMD) scheme. Specifically, we consider the CCN architecture. The model can be used to compare the performance of different multipath forwarding strategies, understanding the reasons why a strategy is better than another one, and to help devising new strategies. We assess the validity of the model by means of simulations.
We also propose a \textbf{new multipath forwarding strategy}, named Fast Pipeline Filling, and we compare it with literature solutions \cite{udugama} \cite{carofigliooptimal}. Moreover, we carry out an experimental campaign by using the \textbf{PlanetLab test-bed} to evaluate a \textbf{CCNx-based implementation} of both our strategy and literature strategies in realistic network settings. All the software is freely available to allow other researchers to reproduce our results \cite{software}.


\begin{figure}[t]
\centering
\includegraphics[scale=0.35]{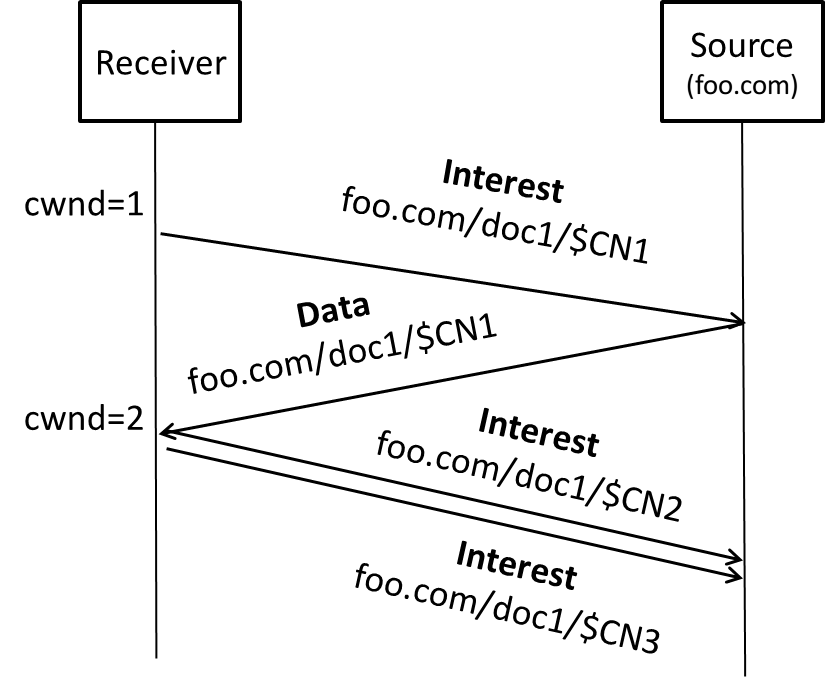}
\caption{Receiver-driven flow control}
\label{f:flow}
\vspace{-10pt}
\end{figure}

\section{Background and related works}
\subsection*{Content Centric Network - CCN}
A CCN addresses contents by using unique hierarchical names \cite{jacobson2009networking} (e.g. foo.com/doc1). Big contents are split into chunks, uniquely addressed by names that include the content name and the chunk number (e.g. foo.com/doc1/\$CNx).
To fetch a chunk, a receiver sends out an Interest message, which includes the chunk name. CCN nodes use a name-based Forwarding Information Base (FIB) to route-by-name Interest messages by using a prefix match logic. A FIB entry contains a name prefix (e.g. foo.com) and a list of \emph{upstream} (inter)faces on which the Interest message can be forwarded towards available sources. When the upstream list contains more than one face, a multipath forwarding strategy singles out a forwarding face, or a set of them e.g. if replication is needed.
During the Interest forwarding process, a CCN node leaves reverse path information  \textless chunk name, \emph{downstream} faces\textgreater in a Pending Interest Table (PIT). When an Interest reaches a node having the requested chunk, the node sends back the chunk within a Data message, which is routed on the downstream path by consuming the information previously left in the PITs. Traversed CCN nodes cache forwarded Data messages, so providing in-network caching functionality.

To download a whole content, a receiver fetches all the related chunks by sending out a sequence of Interest messages. For flow control purposes, the receiver exploits a receiver-driven approach, which consists in limiting the number of in-flight Interests through a congestion window (cwnd). The cwnd size may be constant or regulated by an Additive Increase Multiplicative Decrease (AIMD) control mechanism. For instance, in fig. \ref{f:flow} the receiver initially has cwnd=1 and sends out an Interest message for the first chunk (foo.com/doc1/\$CN1). At the reception of the related Data, the AIMD algorithm sets cwnd=2 and the receiver sends out two Interests for the next two chunks.

CCNx \cite{ccnx} is a Linux-based CCN implementation whose faces are UDP or TCP tunnels. The default multipath strategy implemented in the ccnd daemon (at least up to version 0.8.1) selects the fastest responding face, and performs experiments to determine if other faces can provide faster response. A similar approach of best face selection, but with face ranking mainly based on data loss is proposed in \cite{yi2013case}. From a practical point of view, in these cases only one path is used to fetch a given content; conversely, in this paper we focus on strategies that concurrently use all available paths. The default control mechanism, provided by the ccngetfile application, uses a constant congestion window.

\subsection*{Multipath scenarios}

\begin{figure}[t]
\centering
\includegraphics[scale=0.35]{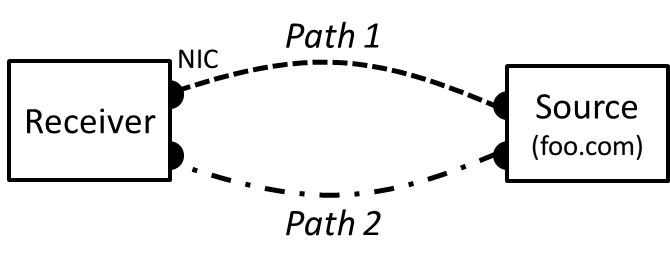}
\caption{End-to-end multipath with multi-homed devices}
\label{f:scen1}
\vspace{-10pt}
\end{figure}

\begin{figure}[t]
\centering
\includegraphics[scale=0.35]{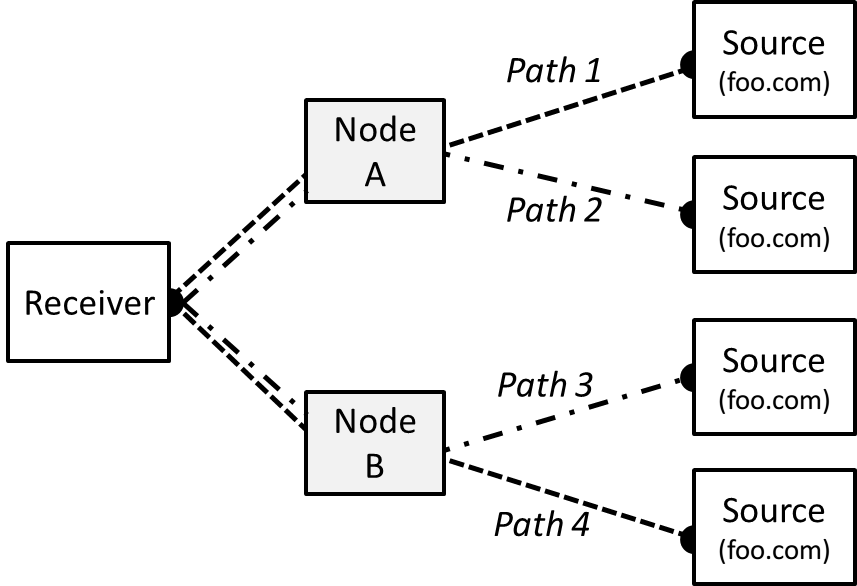}
\caption{End-to-end multipath with server pooling}
\label{f:scen2}
\vspace{-10pt}
\end{figure}

\begin{figure}[t]
\centering
\includegraphics[scale=0.35]{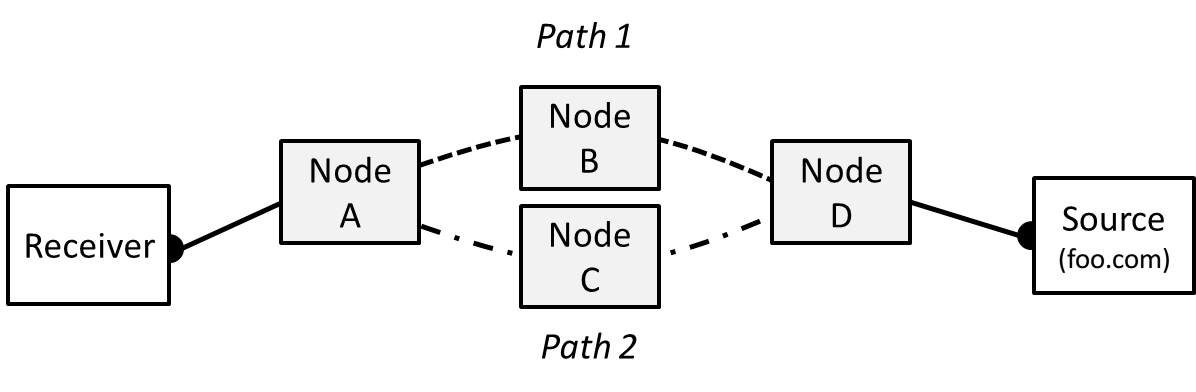}
\caption{In-network multipath}
\label{f:scen3}
\vspace{-10pt}
\end{figure}

\begin{figure}[t]
\centering
\includegraphics[scale=0.35]{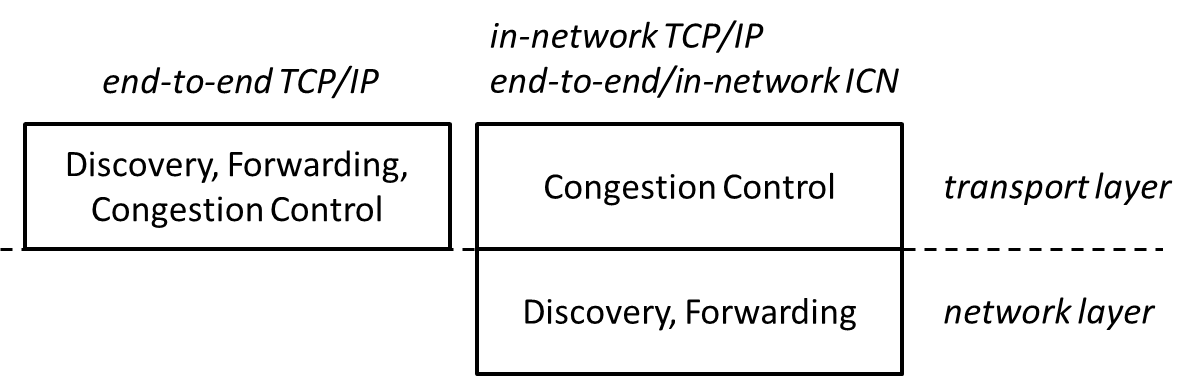}
\caption{Protocol layer of multipath functionality}
\label{f:stack}
\vspace{-10pt}
\end{figure}

It is possible to exploit different network paths to transfer a content in two cases: when there are multiple end-points; when there are multiple network paths between two end-points. We refer to the first case as \emph{end-to-end multipath}, and to the second one as \emph{in-network multipath}.
Figures \ref{f:scen1} and \ref{f:scen2} describe two possible scenarios of end-to-end multipath. In the former ``multi-homing'' case (Fig. \ref{f:scen1}) a receiver and/or a source expose multiple end-points, since they have multiple Network Interface Cards (NICs). 
Different connections among NIC couples can be set up, and these connections are the multiple (overlay) paths.
The second ``server pooling'' case (Fig. \ref{f:scen2}) is characterized by a pool of sources that provide the same content. A receiver, or a middlebox, can set up multiple connections with these sources to fetch a content.
Figure \ref{f:scen3} shows a possible scenario of in-network multipath. The receiver and the source can exchange data through two network paths. In the following subsections, we discuss both the TCP/IP and the ICN multipath approaches to exploit end-to-end and in-network multipath.

\subsection*{TCP/IP multipath}
In TCP/IP networks, multipath issues have been abundantly discussed in the literature \cite{he2008toward}. We briefly report some reference approaches to exploit end-to-end and in-network multipath. 

\emph{End-to-end multipath} - The exploitation of end-to-end multipath requires to discover remote NICs and to split data traffic among them. These operations, and congestion control, can be executed above the IP layer. Consequently, as shown in fig. \ref{f:stack}, end-to-end multipath systems run above the IP network layer, at the communication end-points, without affecting network nodes. Usually, a multipath forwarding strategy splits traffic on a per-packet basis, to maximize the transfer rate. For instance, in the ``multi-homed'' scenario (fig. \ref{f:scen1}) the MultiPath TCP (MPTCP) protocol \cite{rfc6824} sets up parallel subflows among couples of NICs discovered through TCP options, uses a TCP-friendly congestion control per-subflow, and schedules traffic on the different sub-flows according to a specific forwarding strategy \cite{singh2012performance,wischik2011design}. In the ``server pooling'' scenario, a receiver can use a BitTorrent approach to concurrently fetch different file pieces from different sources, discovered with Web means. The transfer of each piece is controlled by a TCP connection.

\emph{In-network multipath} - The exploitation of in-network multipath requires to discover internal network paths and to control the forwarding of traffic inside the network. Consequently, the path discovery and multipath forwarding mechanisms must be necessarily executed by the IP network layer \footnote{We are not considering the possibility of using the source routing IP option since practically is not supported}, thus involving network nodes. Instead, congestion control mechanism can remain above IP, at the communication end-points. However, as shown in fig. \ref{f:stack}, this division of multipath functionality in two different layers creates inter operation issues, of which a crucial one is packet reordering. In fact, if a per-packet strategy were used by IP routers, it would cause out of order packet delivery (as different paths may have different delays) and TCP-based congestion control would wrongly reduce the send-rate, even in absence of congestion \cite{lim2003tcp}. Conversely, out-of-order delivery does not occur in the case of per-flow multipath forwarding strategies. Thus, per-flow strategies are the safest approach in TCP/IP networks, to exploit in-network multipath.

\subsection*{ICN multipath}
A user above the ICN network layer is unaware of where a content is. A user can not choose neither a remote NIC nor a network path. It is the ICN network layer that cares to find the NIC of a source, provide the content, select a network path and send back the content to the requesting user. Consequently, in all multipath scenarios, path discovery and multipath forwarding mechanisms must be necessarily executed by the ICN network layer, while congestion control can remain above the ICN network layer, at the communication end-points, as shown in fig. \ref{f:stack}.

Path discovery is carried out by means of a multipath ICN routing protocol, which discovers a set of alternate paths from nodes to contents, and configures the forwarding tables with multiple output (inter)faces.

Multipath forwarding is carried out through specific strategies that can operate on a per-packet or per-flow (i.e. per-content delivery) basis. To better exploit the transfer capacity of multiple paths, per-packet approaches are currently receiving a greater interest. Moreover, in the CCN architecture, it is easily possible to extend the functionality of the PIT by adding to its main job of reverse routing also the monitoring of path performance parameters, such as number of pending interests or round trip times. These parameters are of undeniable utility to implement multipath strategies. For instance, in \cite{udugama} the authors propose a weighted round robin scheme among faces, whose weights are inversely proportional to the face round trip time; in \cite{carofigliooptimal} the weights are inversely proportional to the number of pending Interest messages; this strategy is claimed to be optimal to maximize user throughput and minimize overall network cost, in case of delay-based congestion control schemes (e.g. TCP Vegas).

Congestion control is an open ICN issue and there is not yet a ``standard'' protocol. Out of delivery may frequently happen in ICN, due to in-network caching and per-packet multipath forwarding strategies. Thus, recent works suggest to use receiver-driven congestion control schemes that do not consider out of order delivery as a  symptom of congestion, but rather infer congestion from other parameters such as increasing delay (delay-based congestion control) \cite{carofigliooptimal}\cite{carofigliomultipath}, and packet loss (loss-based congestion control) \cite{braun2013empirical} \cite{saino2013cctcp}. However, it is not clear, yet, which is the best indicator for congestion. In \cite{prasad2004effectiveness} the authors raise concerns about delay-based approaches due to the small correlation between increased delays (or RTTs) and congestion-related losses in wired Internet measurements. Conversely, it is well-known that loss-based congestion control dramatically suffers from the random packet loss of wireless environments \cite{balakrishnan1997comparison}. In any case, the receiver-driven and connectionless nature of ICN congestion control enables receivers to select the best approach, depending on actual conditions.

Finally, it is useful to note that if multipath may increase the receiving rate, it may also reduce effectiveness of in-network caching, due to the possible routing of the same request on different paths \cite{udugama2013analytical}\cite{rossi2011caching}.


\section{Model of the AIMD receive-rate with multipath forwarding}
We propose an analytical model of the receive-rate of a loss-based AIMD congestion control in presence of a \emph{generic} per-packet multipath forwarding strategy. The model exploits some simplifying approximations that, however, do not impair its accuracy, as we will show in section \ref{s:Analytical and simulation results}.
\begin{figure}[t]
\centering
\includegraphics[scale=0.35]{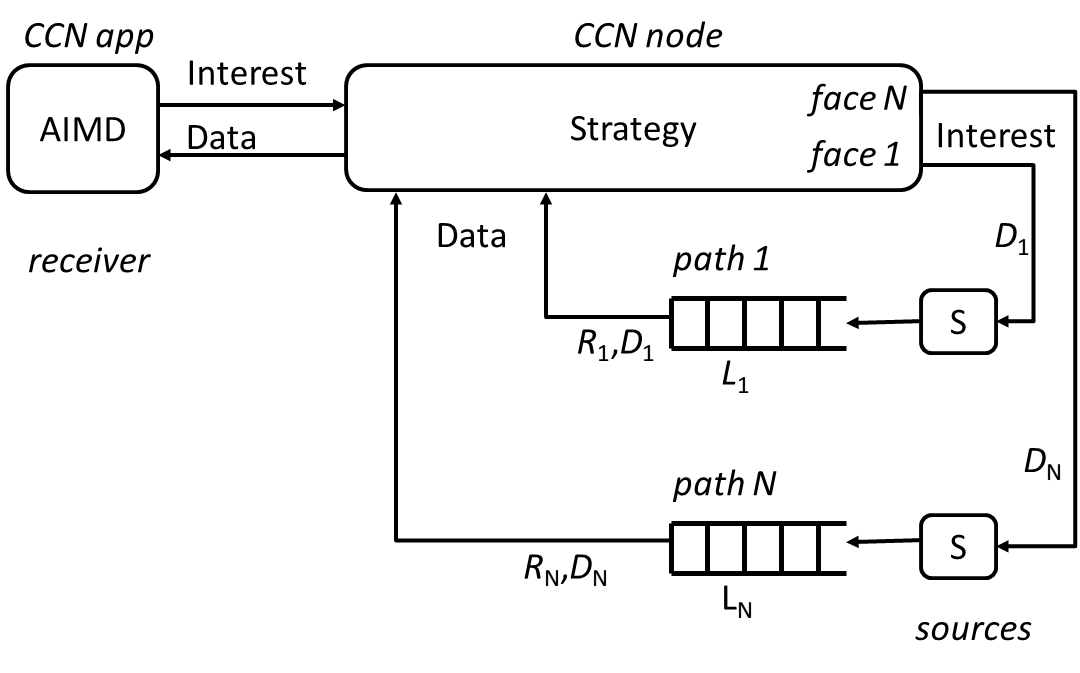}
\caption{Reference network model}
\label{f:model1}
\vspace{-10pt}
\end{figure}

Figure \ref{f:model1} depicts the network model that we use for our analysis: a CCN application (e.g. ccngetfile \cite{ccnx}) is used to fetch a content, and integrates a loss-based AIMD congestion control. The Interest messages generated by the AIMD entity are sent to the underlaying CCN node multipath forwarding functionality, which implements a per-packet strategy. The strategy determines how to distribute the Interest messages on the $N$ upstream paths. When an Interest reaches a source S (repository or in-network cache) at the end of the path, the source sends back the related Data message. The Data message retraces the path followed by the Interest message, in the downstream direction, reaches the CCN node and then the AIMD entity. To simplify the model, we assume that a source (be it a repository or an in-network cache) has all the content chunks, i.e. we are not modeling the case of in-network caches having a subset of content chunks only. 

We also assume that congestion can occur on the downstream path only. Thus, we model the $i$th upstream path as a simple delay line, with a constant propagation delay equal to $D_i$ seconds. In addition, similarly to \cite{jacobson1988congestion}, we model the $i$th downstream path with a fixed propagation delay equal to $D_i$ seconds, with a FIFO buffer of size $L_i$ Data packets, which is emptied with a rate of $R_i$ Data messages per second. This simple queuing system is used to model the slowest link of the downstream path (its bottleneck), while the remaining links of the path are considered lossless.


We consider a receiver-driven congestion control mechanism that always maintains in the network $W$ in-flight Interest messages, related to the next $W$ missing chunks. A new Interest is immediately sent out after the reception of each Data, irrespective of whether the received Data is in order or not. The congestion window size $W$ is controlled by an AIMD scheme, which increases it by one Interest per window $W$ of Data received, and halves the window when a Data loss occurs. We assume that Data loss is detected immediately after the loss occurrence, i.e. we are not considering the possible detection delay e.g. due to time-out overestimation.    

Let us consider the download of a given content. We consider a generic multipath forwarding strategy and define $P_i(H)$ as the \emph{average} number of pending Interest messages injected by the considered strategy on the upstream face $i$, for $i=1 \dots N$, when the strategy is handling a fixed number of $H$ pending Interest of the considered content. We define the vector function $P(H)$ whose elements are $P_i(H)$.

\begin{equation}
\begin{aligned}
& P(H) \longrightarrow \{ P_i(H) \} \\
& \text{s.t.} \\
&\sum P_i(H) = H \\
&P_i(H) \geq 0 \\
&i=1 \dots N \\
\label{eq:0}
\end{aligned}
\vspace{-12pt}
\end{equation}

In the following we will refer to  $P(H)$ as the \emph{sharing function} of the forwarding strategy. This characterization of a strategy is a key feature of our model, which makes it simple and general. In the following description of the model we consider $P(H)$ as a generic function; then in section \ref{s:ssb} we specialize the function $P(H)$ to specific strategies, to evaluate their performance.
To make an example here, the sharing function of a strategy aimed at balancing the number of pending Interest messages across the upstreams faces is $P_i(H) = H/N$, for $i=1 \dots N$.


\begin{figure}[t]
\centering
\includegraphics[scale=0.35]{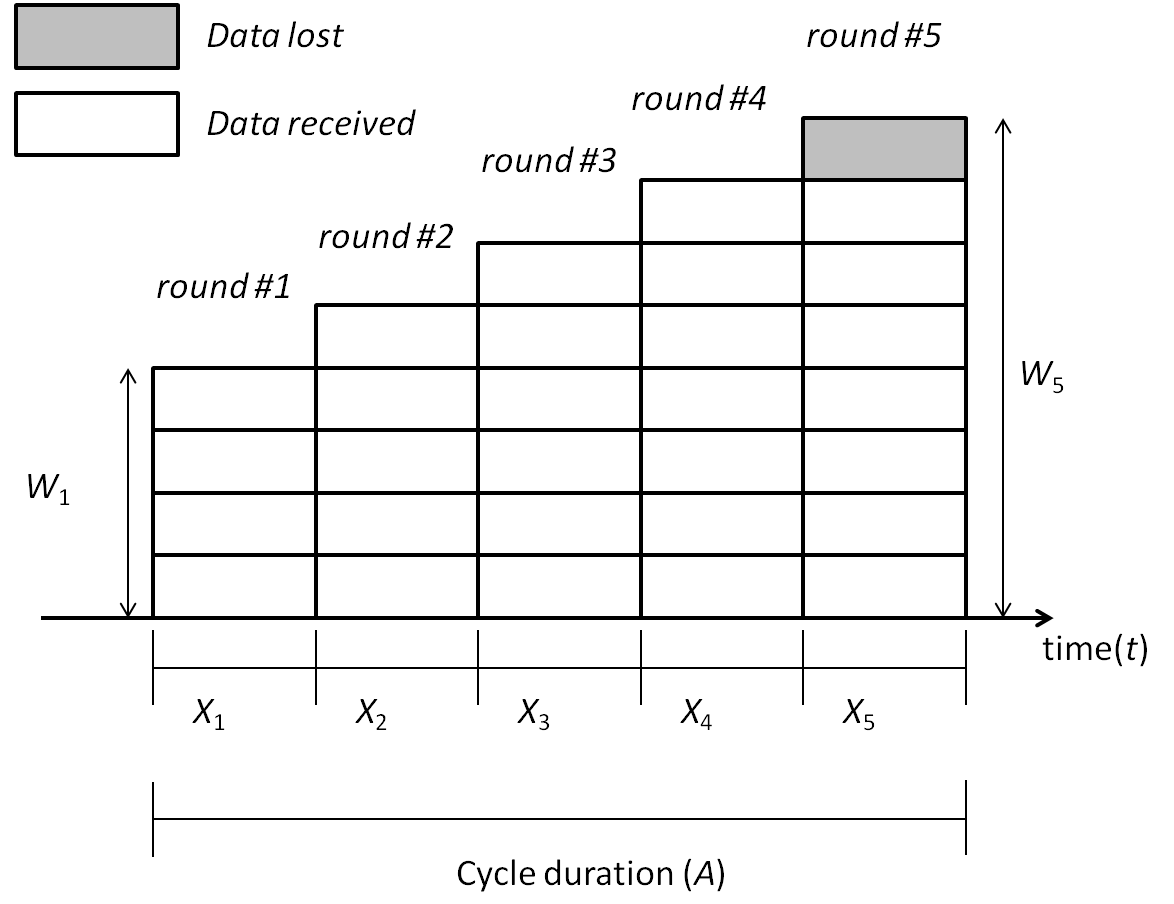}
\caption{Evolution of the AIMD congestion window}
\label{f:model2}
\vspace{-10pt}
\end{figure}

In the considered network model (fig. \ref{f:model1}), the multipath forwarding strategy handles packets controlled by the AIMD entity, with a one-to-one relationship, thus the overall number of pending Interest $H$ handled by the strategy is equal to the congestion window size $W$ of the AIMD.

We model the evolution of the congestion window size in terms of ``rounds''. A round starts when the AIMD algorithm changes the value of the window size $W$ and ends either when $W$ Data messages are received or when they would have been received if loss did not happen (i.e. at the expiry of the ideal time-out related to the expected receipt of such messages). Thus, the window size remains constant during a round.

We define as ``cycle'' a sequence of rounds without losses following a lossy round and including the first following lossy round. 
For instance, in fig. \ref{f:model2} we have a cycle made up of five rounds. In the first round the congestion window $W_1$ is equal to 4 Interest messages. In the fifth round a Data loss occurs; thus, this is the last round of the cycle and the maximum congestion window reached during the cycle is $W_5=8$.

Since the network model does not consider random phenomena (e.g. random loss, delay, etc.), the congestion window behavior is periodic and formed by a sequence of equal cycles. Therefore, to evaluate average performance it is sufficient to compute average performance in a cycle.

To evaluate the average receive-rate $Y$ in a cycle, we approximate the number of received Data messages with the number of sent Data messages, which is equal to the number of sent Interest messages $T$.
Defining with $A$ the duration of a cycle, the receive-rate $Y$ can be written as:

\begin{equation}
Y = \frac{T}{A}
\label{eq:1}
\end{equation}

Let us now determine the value of $T$. During a cycle, the congestion window increases from a minimum value equal to $\lfloor W_{max} / 2 \rfloor$ up to the maximum value $W_{max}$. The number $T$ of Interest messages sent out by the AIMD entity in a cycle can be written as:
\begin{equation}
T = \sum_{k=\lfloor W_{max}/2 \rfloor}^{W_{max}} k
\label{eq:2}
\end{equation}
In a round, a Data loss occurs when the number of pending Interest injected in any path by the strategy is greater than the \emph{pipeline capacity} of the path, i.e. the sum of the bandwidth-delay product $R_i \cdot (2 \cdot D_i)$ and of the buffer space ($L_i$). Therefore, the maximum congestion window $W_{max}$ reached in a round can be evaluated by solving the following integer maximization:
\begin{equation}
\begin{aligned}
& \text{max}
&& W \\
& \text{s.t.}
&& P_i(W) \leq R_i \cdot (2 \cdot D_i)+L_i, \; i = 1, \ldots, N.
\end{aligned}
\end{equation}

Let us now determine the duration $A$ of a cycle. As shown in fig. \ref{f:model2}, $A$ is equal to the sum of the duration $X_k$ of the rounds of the cycle, where $k$ is the round index within the cycle, i.e.:
\begin{equation}
A = \sum_{k=1}^{W_{max}-\lfloor W_{max}/2 \rfloor} X_k
\label{eq:3}
\end{equation}
A round $k$ lasts for the time needed to exchange a number of Data messages equal to the congestion window $W_k$ of the AIMD during that round, which is equal to
\begin{equation}
W_k = \lfloor W_{max}/2 \rfloor + (k-1)
\end{equation}
Defining as $B_k$ the overall receive-rate in the round $k$, the duration $X_k$ of round $k$ can be written as:
\begin{equation}
X_k = \frac{W_k}{B_k}
\end{equation}
Each path contributes to $B_k$. The contribution $B_{k,i}$ of the $i$th path is equal to the ratio between the number of in-flight Interest messages $P_i(W_k)$ on the path and the path round trip time $RTT_i$. Thus, we can write:

\begin{equation}
B_k = \sum_{i=1}^{N} B_{k,i}
\end{equation}
\begin{equation}
B_{k,i} = \frac{P_i(W_k)}{RTT_i}
\label{eq:bki}
\end{equation}
\begin{equation}
RTT_i = max\{2 \cdot D_i, P_i(W_k)/R_i\}
\label{eq:rtt}
\end{equation}
The above equation gives an approximation of $RTT_i$ similar to the one used in \cite{jacobson1988congestion}. Indeed, when the number of in-flights  messages is lower than the bandwidth-delay product, the path performance are \emph{delay-dominated} and $RTT_i$ is equal to the propagation delay. Otherwise, the path performance are \emph{bandwidth-dominated} and $RTT_i$ is equal to the ratio between the number of in-flight message and the available rate $R_i$.

\section{Multipath forwarding strategies}
\label{s:ssb}
In this section we present five strategies: two rather generic ones, namely Pending Interest Equalization (PE) and RTT Equalization (RE); the strategy proposed in \cite{udugama} (UG); the strategy proposed in \cite{carofigliooptimal} (CF); and our own, Fast Pipeline Filling (FPF). For each strategy we model the sharing function $P(H)$ (eq. \ref{eq:0}), which enables to analytically compute the receive-rate $Y$ by means of eq. \ref{eq:1}. These strategies monitor the characteristics of the path (e.g. number of pending Interest, RTT, etc.) for each content. 

\subsection{Pending Interest Equalization (PE)}
The goal of this strategy is to balance the number of pending Interest on the different $N$ paths. For each received Interest the strategy chooses the face with the lowest number of pending Interest messages; in case of equality, a random face is chosen. The sharing function $P(H)$ can be readily written as
\begin{equation}
P_i(H) = H/N
\end{equation}


%

%
%

%

\subsection{Round Trip Time Equalization (RE)}
The goal of this strategy is to equalize the round trip time observed on the different faces. For each received Interest, the strategy chooses the face with the lowest RTT. In doing so, the RTTs tend to be equalized since increasing the number of pending Interests on a path, the path RTT increases too or remains constant (see eq. \ref{eq:rtt}). From another point of view, this strategy could also be seen as a greedy approach to minimize the RTT. 

Since the RTT is not a linear function (eq. \ref{eq:rtt}), it is not easy to evaluate the sharing function $P(H)$ of the strategy with a closed formula. For this reason, we resort to the recursive algorithm \ref{a:PiRTTE} below, in which at each step the face with the lowest RTT is selected.


\begin{algorithm}
\caption{Computation of $P(H)$ for RTT Equalization}
\label{a:PiRTTE}
\begin{algorithmic}[1]
\Procedure {RTT}{$d,r,pi$}
\State \textbf{return} $max\{2 \cdot d, pi/r\}$
\EndProcedure
\\
\Procedure {P}{H}
\State $P_i(H) = 0$ for $i = 1 \dots N$
\State $S = 1 \dots N$
\For {$x=1 \dots H$}
\State Select $i \in S$ s.t. $RTT(D_i,R_i,P_i(H))$ is min
\State $P_i(H) = P_i(H) + 1$
\EndFor
\State \textbf{return} $\{P_i(H)\}$
\EndProcedure
\end{algorithmic}
\end{algorithm}

\subsection{Strategy of \cite{udugama} (UG)}
\label{s:ug}
This strategy distributes incoming Interests among faces by using a weighted round robin logic, with the weight $z_i$ of face $i$ being inversely proportional to that face round trip time $RTT_i$, i.e.

\begin{equation}
\label{eq:ziUG}
z_i = \frac{1}{RTT_i \cdot \sum_{j=1}^{N} RTT_j^{-1}}
\end{equation}

Rather surprisingly, we found that the sharing function $P(H)$ of this RTT-based strategy is equal to the one of the pending Interest equalization strategy. Indeed, during a round $k$ the receive-rate $B_{k,i}$ of path $i$ is the one reported in eq. \ref{eq:bki}. Since the UG strategy consists of a weighted round robin scheme, $B_{k,i}$ is also equal to the overall rate $B_k$ multiplied by the weight $z_i$. Consequently we can write the following equations:
\begin{equation}
\begin{aligned}
&\frac{P_i(H)}{RTT_i} = \bigg( \sum_{j=1}^{N} \frac{P_j(H)}{RTT_j} \bigg) \cdot z_i \mbox{    for   } i=1 \dots N \\
& \sum_{i=1}^{N} P_i(H) = H
\label{eq:udu}
\end{aligned}
\end{equation}
The solution of these equations is simply $P_i(H)=P_j(H)=H/N$ for any $i,j$, i.e. the sharing function of the pending Interest equalization (PE) strategy. This implies that the two strategies will result in the same receive-rate, even though the PE strategy has a simpler implementation since it does not require to estimate the RTT.

\subsection{Strategy of \cite{carofigliooptimal} (CF)}
\label{s:caro}
This strategy distributes incoming Interests on faces by using a weighted round robin logic, with the weight $z_i$ of face $i$ being inversely proportional to its number of pending Interest messages $P_i$, i.e.
\begin{equation}
z_i = \frac{1}{P_i \cdot \sum_{j=1}^{N} P_j^{-1}}
\label{eq:ziCF}
\end{equation}


The sharing function can be computed by using eq. \ref{eq:ziCF} in eqs. \ref{eq:udu}. After some simple algebra eqs. \ref{eq:udu} can be written as:

\begin{equation}
\frac{P_i(H)}{\sqrt RTT_i} = \frac{P_j(H)}{\sqrt RTT_j} \mbox{    for any   } i,j
\label{eq:caro}
\end{equation}

\begin{equation}
\sum_{i=1}^{N} P_i(H) = H
\end{equation}

Eq. \ref{eq:caro} shows that the sharing function $P(H)$ of the CF strategy resembles the one of a weighted PE strategy, whose weights are the square root of the round trip times. Thus, paths with higher round trip time will have more pending Interests with respect to the PE strategy.

To evaluate the sharing function $P(H)$, we resort to an approximated method given by the following iterative algorithm \ref{a:PiCF}. At each iteration step, the face with the lowest $P_i(H)/\sqrt RTT_i$ is selected.

\begin{algorithm}
\caption{Computation of $P(H)$ for CF strategy}
\label{a:PiCF}
\begin{algorithmic}[1]
\Procedure {P}{H}
\State $P_i(H) = 0$ for $i = 1 \dots N$
\State $S = 1 \dots N$
\For {$x=1 \dots H$}
\State Select $i \in S$ s.t.
\State $P_i(H) / \sqrt{RTT(D_i,R_i,P_i(H))}$ is min
\State $P_i(H) = P_i(H) + 1$
\EndFor
\State \textbf{return} $\{P_i(H)\}$
\EndProcedure
\end{algorithmic}
\end{algorithm}

\subsection{Fast Pipeline Filling (FPF)}
Our FPF strategy has been motivated by insights enabled by our model. Its goal is to completely fill the pipeline capacity of the different paths, and to achieve this saturation condition as fast as possible. In doing so, the value $W_{max}$ reached by the congestion window during a cycle is the maximum possible one, the cycle duration $A$ is the shortest possible one, and this choice maximizes the receive-rate of eq. \ref{eq:1}.

For each received Interest, the FPF strategy identifies the set $S$ of faces whose number of pending Interest messages is lower than the related pipeline capacity ($C_i$). Within this set, the strategy selects the face with the lowest RTT. The sharing function $P(H)$ can be computed by means of algorithm \ref{a:PiFPF}.

\begin{algorithm}
\caption{Computation of $P(H)$ for FPF }
\label{a:PiFPF}
\begin{algorithmic}[1]
\Procedure {P}{H}
\State $P_i(H) = 0$ for $i = 1 \dots N$
\State $C_i = 2 \cdot R_i \cdot D_i+L_i$
\For {$x=1 \dots H$}
\State Form the set $S$ of face indexes $i$ s.t. $P_i(H) < C_i$
\State Select $i \in S$ s.t. $RTT(D_i,R_i,P_i(H))$ is min
\State $P_i(H) = P_i(H) + 1$
\EndFor
\State \textbf{return} $\{P_i(H)\}$
\EndProcedure
\end{algorithmic}
\end{algorithm}

\section{Analytical and simulation results}
\label{s:Analytical and simulation results}
To assess the validity of the analytical model, we developed an event-driven Matlab simulator reproducing the scenario reported in fig. \ref{f:model1}; then we carried out a set of tests considering two paths. The first path has a delay $D_1=20ms$, a queue length $L_1 = 20$ Data messages and a transmission rate $R_1 = 10$ Mbps. The length of a Data message is 4876 bytes, 4096 bytes of payload and 780 bytes of CCN/UDP/IP overhead. This value has been taken from CCNx measurements. Then, we varied one at a time the delay $D_2$ and the transmission rate $R_2$ of the second path, while keeping the unvaried parameter equal to the one of path 1.
We compared the receive-rates of the considered multipath forwarding strategies. We point out that the aim of the comparison is not to judge which is the best strategy, considering also that our scenario is very simple and these strategies do not have all the same objectives. The aim of the comparison is to show how the model can be used to gain insights on the behavior of the receive-rate in presence of a multipath forwarding strategy.        

\begin{figure}[t]
\centering
\includegraphics[scale=0.50]{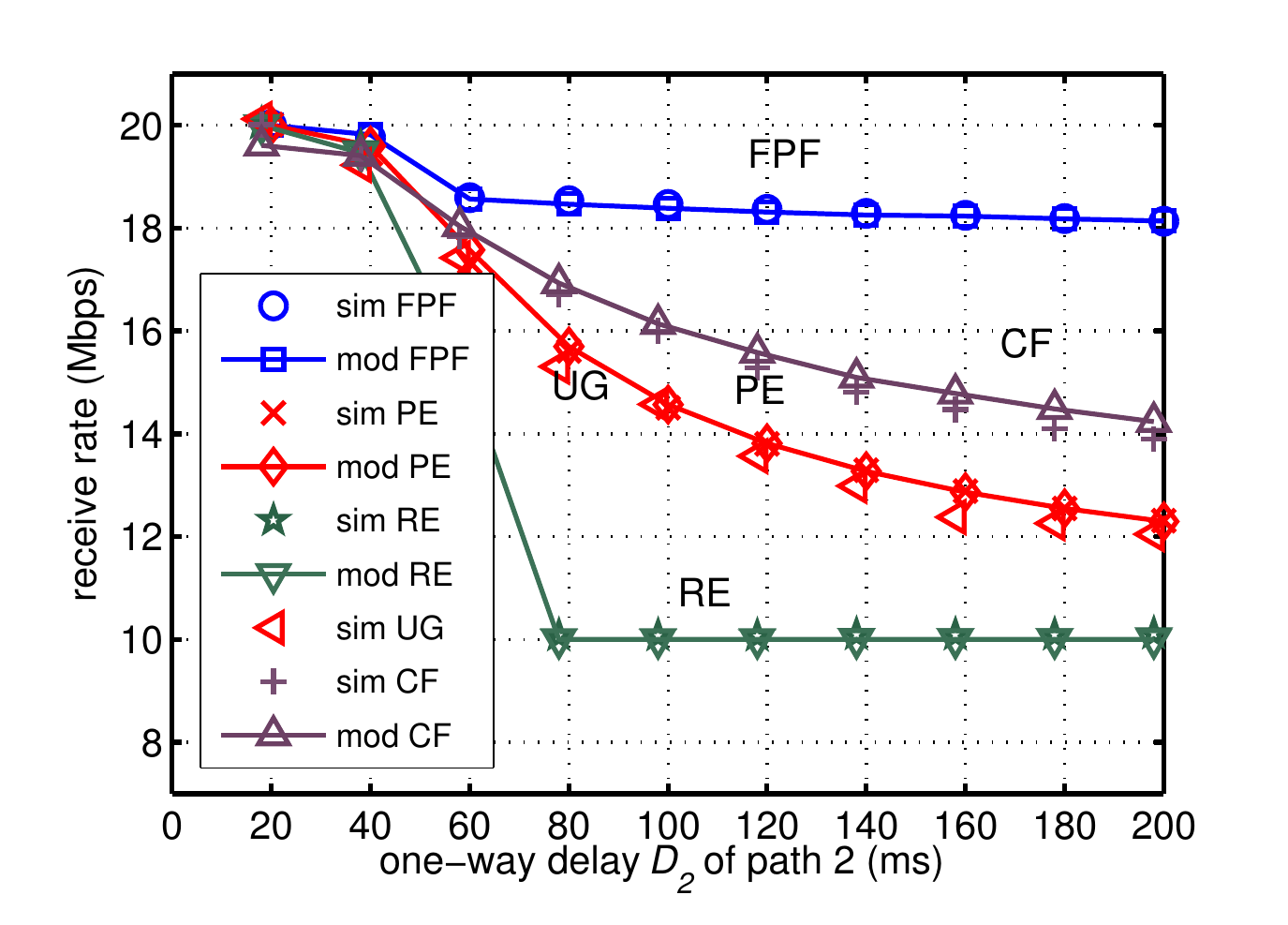}
\caption{Receive-rate $Y$ for two paths versus $D_2$, with $D_1=20$ ms, $R_1=R_2=10$ Mbps, $L_1=L_2=20$ Data messages}
\label{f:vsdelay}
\vspace{-10pt}
\end{figure}

Fig. \ref{f:vsdelay} reports the receive-rate versus the delay of the second path. We observe that model (mod) and simulation (sim) results are very close to each other, thus confirming the validity of the model in the scenario of fig. \ref{f:model1}.

In case of homogeneous paths ($D_2=D_1=20 ms$) all strategies provide the same performance. In average, they equally share the load on both paths, and this is an optimal result for the receive-rate, in case of symmetric paths.

\begin{figure}[t]
\centering
\includegraphics[scale=0.50]{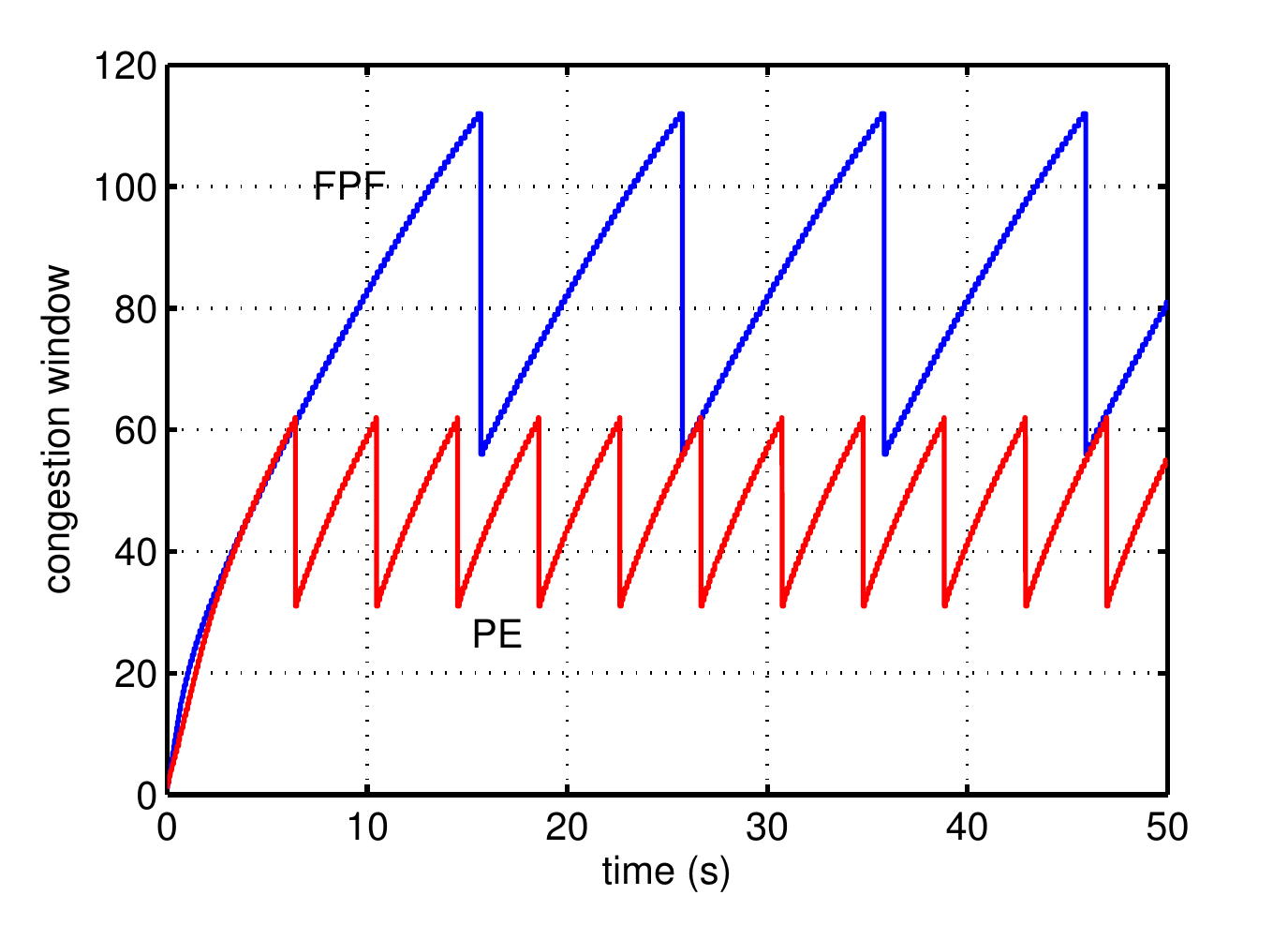}
\caption{Congestion window versus time in case of two paths, with $D_1=20$ ms, $D_2=120$ ms, $R_1=R_2=10$ Mbps, $L_1=L_2=20$ Data messages}
\label{f:window}
\vspace{-10pt}
\end{figure}

As the delay $D_2$ increases, the FPF strategy shows the best performance, since it is able to quickly fill the capacity of the pipelines of both paths. As a consequence the AIMD congestion window reaches the highest possible value between data loss events and the receive-rate performance is the best one. This behavior is shown in fig. \ref{f:window}, which reports the evolution of the congestion window for $D_2=120$ ms, for the PE and FPF strategies. The capacity of the pipeline of paths 1 and 2 is $C_1 = 2 \cdot R_1 \cdot D_1\approx 30$ and $C_2 \approx 81$ Data messages, respectively.  As the congestion window increases up to 30, the FPF strategy injects all messages on path 1, which has the lowest RTT. When the window is greater than 30, the FPF strategy maintains the number of in-flight Interests on path 1 equal to 30 and starts to inject additional in-flight Interest messages in path 2. A first loss occurs when the congestion window becomes greater than the sum of the pipeline capacity of the two paths, i.e. $C_1+C_2=111$. After the first loss, the congestion window drops to 55 and then restarts its growth. The FPF strategy maintains path 1 filled with 30 in-flight Interest messages, other messages are injected in path 2 and a new loss occurs when the congestion window reaches again the value 111.

Fig. \ref{f:vsdelay} confirms the findings anticipated in section \ref{s:ug}: the PE and the UG strategies have the same (average) performance. Their sharing function, which equalizes the number of in-flight Interests, makes the smallest pipeline a limiting factor of the AIMD growth. Indeed, when the smallest pipeline is filled with in-flight messages a drop occurs, even if the pipelines of the other paths are partially available. This partial exploitation of pipelines explains the lower performance with respect to the FPF strategy. For instance, in the scenario of fig. \ref{f:window}, as the congestion window increases, the PE strategy equally distributes the in-flight Interest messages between the two paths. Consequently, when the congestion window reaches the value 62,  there are 31 in-flight Interest messages on path 1, this value is above the capacity of the pipeline of path 1 and a first loss occurs. After this first loss, the congestion window drops to 31 and restarts its growth; the PE strategy equally distributes the in-flight Interests between the two paths and a new loss occurs when the congestion window reaches again 62.

Fig. \ref{f:vsdelay} shows that the performance of the CF strategy is in-between the FPF and PE/UG ones. Since the CF strategy behaves as a weighted PE whose weights are the square root of the RTTs (see section \ref{s:caro}), it maintains a greater value of in-flight Interests on path 2 (whose RTT is greater), with respect to the PE/UG strategy. This allows the AIMD to reach a greater value of the congestion window between losses, i.e. to achieve a greater receive-rate. However, loss typically occurs on path 1 before having saturated the pipeline capacity of path 2; for this reason the performance of the CF strategy is lower than that of FPF.

The RE strategy results in the worst performance in terms of receive-rate. As $D_2$ increases, it tends to waste the second path since its RTT is greater than the one of path 1; thus, the receive-rate decreases to the rate of path 1, i.e. 10 Mbps.

\begin{figure}[t]
\centering
\includegraphics[scale=0.50]{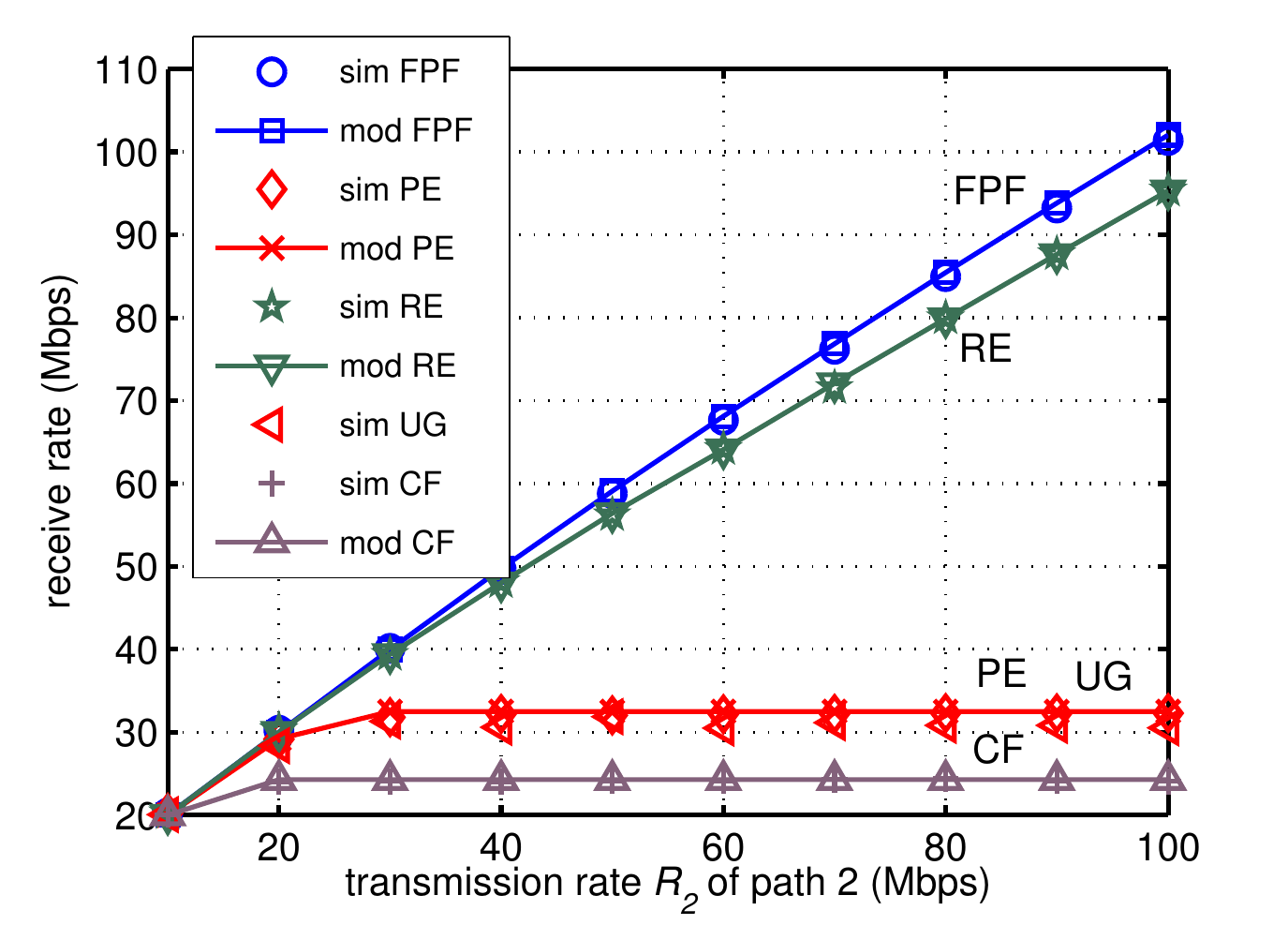}
\caption{Receive-rate $Y$ for two paths versus $R_2$, with $D_1=D_2=20$ ms, $R_1=10$ Mbps, $L_1=L_2=20$ Data messages}
\label{f:vsrate}
\vspace{-10pt}
\end{figure}

Figure \ref{f:vsrate} shows the receive-rate versus the rate $R_2$ of path 2. The FPF strategy provides the best performance. The RE strategy performs rather well since it favors path 2, which has the greater rate and, consequently, a lower RTT, due to its smaller queening delay (see eq. \ref{eq:rtt}). The performance of the PE and UG strategies are limited to the small pipeline capacity of path 1 and the achieved rate is roughly two times the rate of path 1, i.e. 20 Mbps. For the CF strategy this scenario is clearly critical, since it tends to use the slower path 1 even more than the PE and UG approaches, since path 1 has an higher RTT.

\begin{figure}[t]
\centering
\includegraphics[scale=0.35]{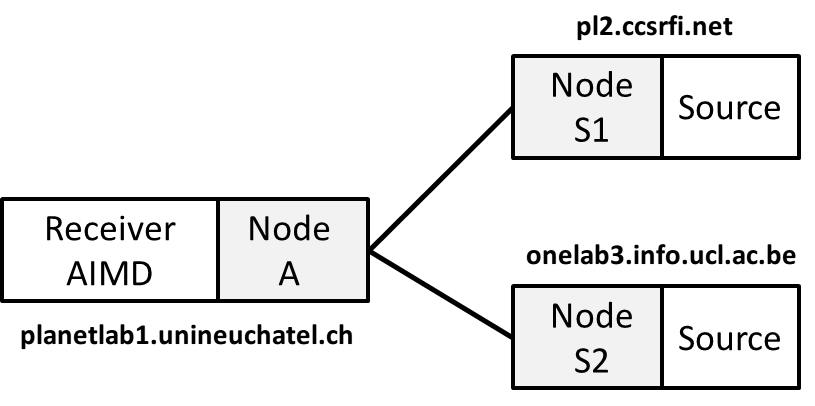}
\caption{PlanetLab scenario 1}
\label{f:plscen1}
\vspace{-10pt}
\end{figure}

\begin{figure}[t]
\centering
\includegraphics[scale=0.35]{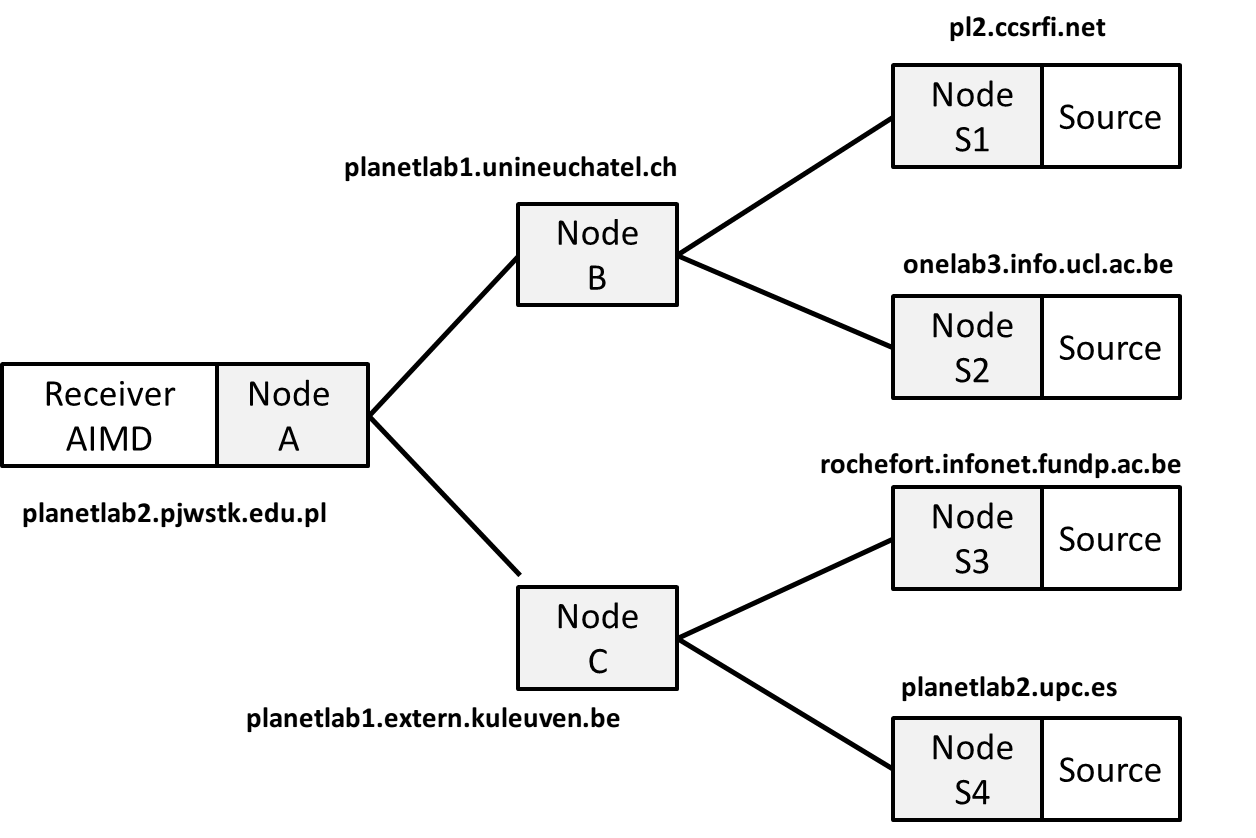}
\caption{PlanetLab scenario 2}
\label{f:plscen2}
\vspace{-10pt}
\end{figure}

\begin{table}
    \centering
    \caption{UDP saturation rate (Mbps) and RTT (ms)}
    \label{table:1}
    \begin{tabular}{l c c c}
    Scenario & \# Link & \# Rate & RTT \\ 
    \hline
    \hline
    1 & A-S1  & 4.8 & 23.1 \\ 
    1 & A-S2  & 8.5 & 17.7 \\ 
    2 & A-B  & 8.2 & 39.2 \\ 
    2 & A-C  & 9.8 & 27.1 \\
    2 & B-S1  & 4.8 & 23.1 \\ 
    2 & B-S2  & 8.5 & 17.7 \\ 
    2 & C-S3  & 7.3 & 17.1 \\ 
    2 & C-S4  & 7.5 & 62.3 \\
    
    \end{tabular}
    \vspace{-10pt}
\end{table}

\begin{figure}[t]
\centering
\includegraphics[scale=0.50]{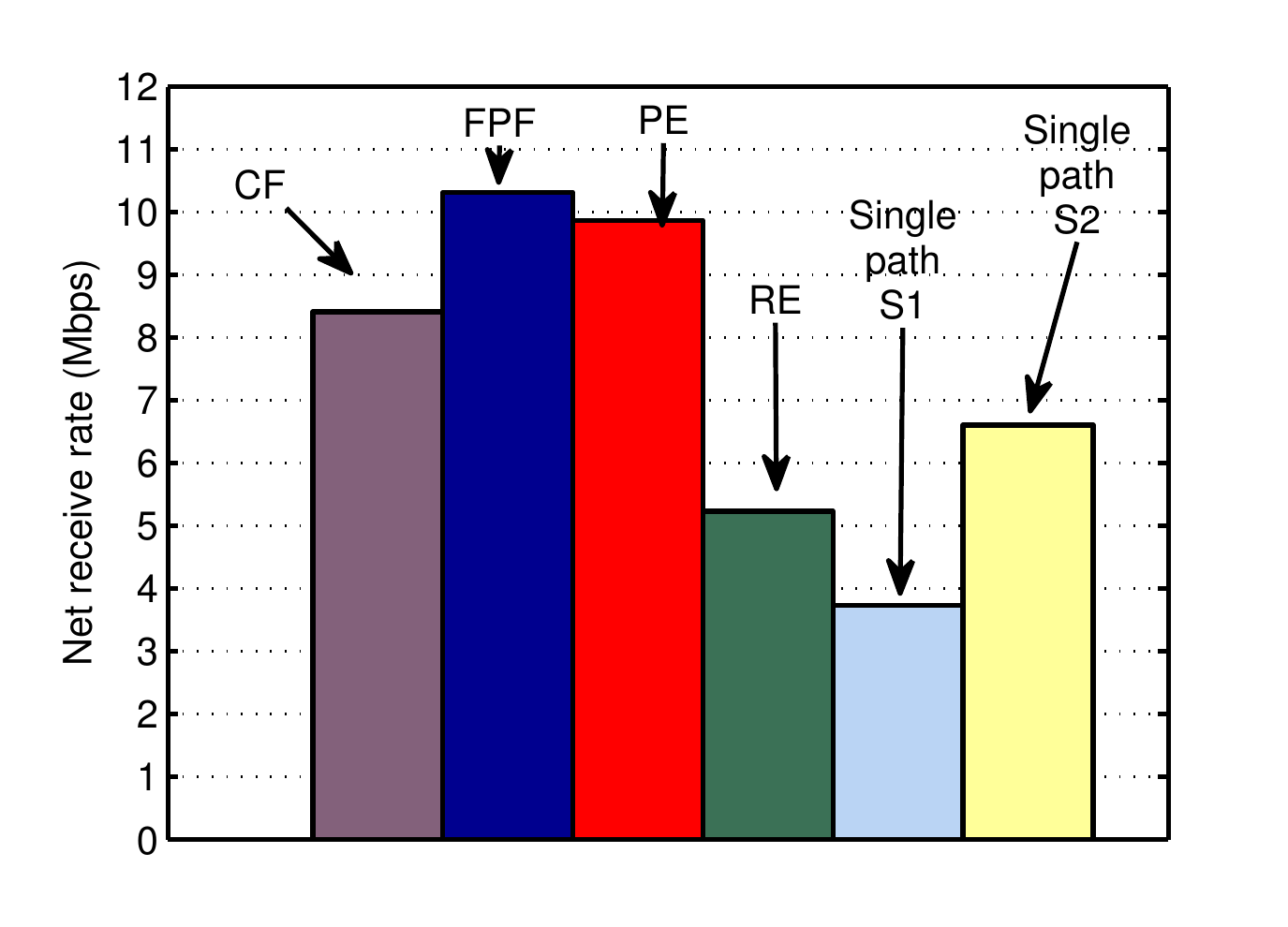}
\caption{Net receive-rate for PlanetLab scenario 1}
\label{f:plrate1}
\vspace{-10pt}
\end{figure}

\begin{figure}[t]
\centering
\includegraphics[scale=0.50]{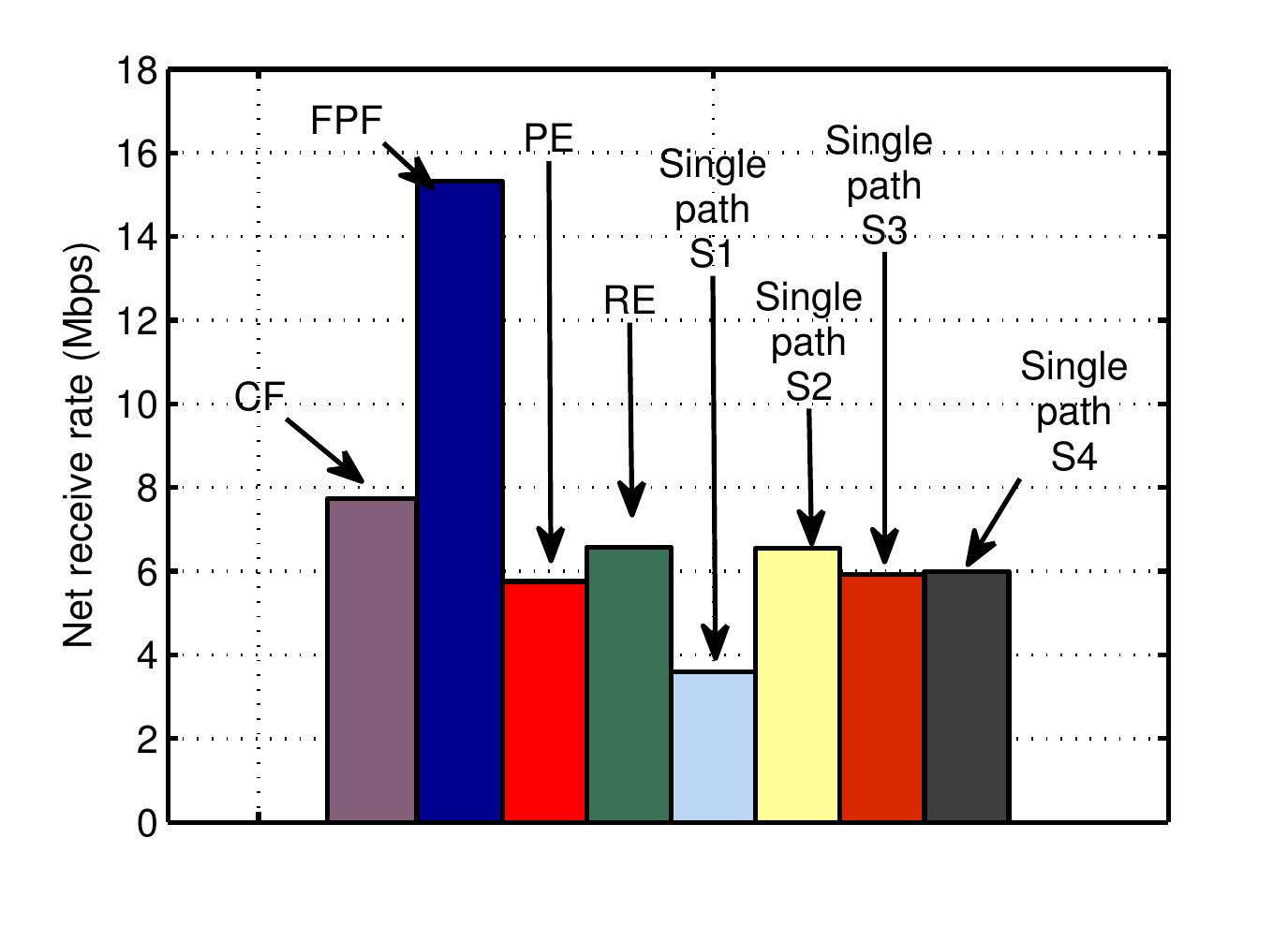}
\caption{Net receive-rate for Planetlab scenario 2}
\label{f:plrate2}
\vspace{-10pt}
\end{figure}

\section{PlanetLab testbed results}
To derive a simple and general receive-rate analytical model we used a simple network scenario (fig. \ref{f:model1}). In this section we want to verify if the findings of our model can be used to \emph{qualitatively} explain the behavior of the considered strategies also in more complex network scenarios, including cascades of CCN nodes.     
We implemented the FPF, RE, PE and CF multipath forwarding strategies and the AIMD congestion control in CCNx (see Appendix I for some details about the implementation). We evaluated the performance in the two scenarios reported in figs. \ref{f:plscen1} and \ref{f:plscen2}, whose nodes are PlanetLab devices. The source nodes S1, S2, S3, and S4 are CCN repositories, which offer the same file of size 50 Mbytes to a receiver. For each scenario we performed 30 measurement batches. A batch consists of a multipath download by using the FPF, RE, PE and CF strategies one at a time, and a single-path download from all sources one at a time. These batches provided measurements whose 95\% confidence interval is below the 5\% of the average value computed among the batches. 

Table \ref{table:1} reports the UDP saturation rate measured with the Iperf tool and the RTT of the involved Internet overlay links.

Figure \ref{f:plrate1} shows the net receive-rate for the PlanetLab scenario 1, i.e. the receive-rate without the CCN/UDP/IP overhead. The FPF strategy provides the highest receive-rate, quite close to the sum of the receive-rates of all single paths. The difference between the single path rates of fig. \ref{f:plrate1} and the UDP saturation rates shown in Tab. \ref{table:1} is mainly due to the CCN overhead and to the oscillatory behavior of the AIMD control. The PE strategy provides a good result too, given that paths have similar pipeline capacities; in this case the equalization of the number of pending Interest is a good choice, as done by the PE and FPF strategies. Other strategies unbalance the number of pending Interests and provide worse performance. The RE strategy tends to prefer the path from R1, which has a lower RTT (i.e. about 17 ms versus the 23 ms of the path from R2). Conversely, the CF strategy tends to prefer the path from R2, due to its higher RTT.

Figure \ref{f:plrate2} reports the net receive-rate for the PlanetLab scenario 2. This is a more heterogeneous scenario, in which the multiple paths have rather different values of the delay and transfer rates (see Tab. \ref{table:1}). All the considered strategies suffer of this heterogeneity, with the exclusion of our FPF. It is noteworthy that the receive-rate provided by the FPF strategy does not reach the sum of the four single-paths rates, since the capacities of links A-B and A-C limit the full exploitation of the following links towards the sources.

\section{Conclusion}
Multipath plays an important role in ICN and the analytical model proposed in this paper may be useful to understand the performance of multipath forwarding strategies as a function of key parameters such as delay, transfer rate and queuing space of available paths. The model has highlighted that a good forwarding strategy \textit{to maximize the receive-rate} should control the pending Interests injected in the different paths so as to fill the capacity of the pipelines of such paths. This is the rationale followed by the FPF strategy, which achieved the best performance in analytical, simulation and PlanetLab results.

The proposed model is useful to understand the receive-rate behavior of forwarding strategies in fixed networks, where packet losses are usually due to congestion. However, in wireless networks losses may frequently occur for other reasons and this is not taken into account by our current model, as well as by the FPF strategy. 
Future work will be aimed to consider random losses in the network model of fig. \ref{f:model1}, devise a multipath forwarding strategy suitable also for wireless environments, and consider other performance maximization figures, e.g. delay.


\bibliographystyle{abbrv}
\bibliography{paper}

\begin{thebibliography}{10}

\bibitem{ccnx}
{CCN}x project.
\newblock http://www.ccnx.org.

\bibitem{software}
M. path sw.
\newblock https://www.dropbox.com/s/hoi9lyi8yyq6j8o/MPSoftware.zip.

\bibitem{balakrishnan1997comparison}
H.~Balakrishnan, V.~N. Padmanabhan, S.~Seshan, and R.~H. Katz.
\newblock A comparison of mechanisms for improving tcp performance over
  wireless links.
\newblock {\em Networking, IEEE/ACM Transactions on}, 5(6):756--769, 1997.

\bibitem{braun2013empirical}
S.~Braun, M.~Monti, M.~Sifalakis, and C.~Tschudin.
\newblock An empirical study of receiver-based aimd flow-control strategies for
  ccn.
\newblock In {\em Computer Communications and Networks (ICCCN), 2013 22nd
  International Conference on}. IEEE, 2013.

\bibitem{carofigliomultipath}
G.~Carofiglio, M.~Gallo, L.~Muscariello, and M.~Papalini.
\newblock Multipath congestion control in content-centric networks.
\newblock 2013.

\bibitem{carofigliooptimal}
G.~Carofiglio, M.~Gallo, L.~Muscariello, M.~Papalini, and S.~Wang.
\newblock Optimal multipath congestion control and request forwarding in
  information-centric networks.
\newblock In {\em Network Protocols (ICNP), 2013 21st IEEE International
  Conference on}, 2013.

\bibitem{rfc6824}
A.~Ford, C.~Raiciu, M.~Handley, and O.~Bonaventure.
\newblock {TCP Extensions for Multipath Operation with Multiple Addresses}.
\newblock RFC 6824 (Experimental), Jan. 2013.

\bibitem{he2008toward}
J.~He and J.~Rexford.
\newblock Toward internet-wide multipath routing.
\newblock {\em Network, IEEE}, 22(2):16--21, 2008.

\bibitem{jacobson1988congestion}
V.~Jacobson.
\newblock Congestion avoidance and control.
\newblock 18(4):314--329, 1988.

\bibitem{jacobson2009networking}
V.~Jacobson, D.~K. Smetters, J.~D. Thornton, M.~F. Plass, N.~H. Briggs, and
  R.~L. Braynard.
\newblock Networking named content.
\newblock In {\em Proceedings of the 5th international conference on Emerging
  networking experiments and technologies}. ACM, 2009.

\bibitem{lim2003tcp}
H.~Lim, K.~Xu, and M.~Gerla.
\newblock Tcp performance over multipath routing in mobile ad hoc networks.
\newblock In {\em Communications, 2003. ICC'03. IEEE International Conference
  on}. IEEE, 2003.

\bibitem{prasad2004effectiveness}
R.~S. Prasad, M.~Jain, and C.~Dovrolis.
\newblock On the effectiveness of delay-based congestion avoidance.
\newblock In {\em Proc. PFLDNet}, 2004.

\bibitem{rossi2011caching}
D.~Rossi and G.~Rossini.
\newblock Caching performance of content centric networks under multi-path
  routing (and more).
\newblock {\em Relat{\'o}rio t{\'e}cnico, Telecom ParisTech}, 2011.

\bibitem{saino2013cctcp}
L.~Saino, C.~Cocora, and G.~Pavlou.
\newblock Cctcp: A scalable receiver-driven congestion control protocol for
  content centric networking.
\newblock In {\em Communications (ICC), 2013 IEEE International Conference on}.
  IEEE, 2013.

\bibitem{singh2012performance}
A.~Singh, C.~Goerg, A.~Timm-Giel, M.~Scharf, and T.-R. Banniza.
\newblock Performance comparison of scheduling algorithms for multipath
  transfer.
\newblock In {\em Global Communications Conference (GLOBECOM), 2012 IEEE}.
  IEEE, 2012.

\bibitem{udugama}
A.~Udugama and C.~Goerg.
\newblock Adaptation of multi-path forwarding strategies for ccn to operate
  under mobility.
\newblock In {\em Presented at CCNxCon2013}. Parc, 2013.

\bibitem{udugama2013analytical}
A.~Udugama, S.~Palipana, and C.~Goerg.
\newblock Analytical characterisation of multi-path content delivery in content
  centric networks.
\newblock In {\em Future Internet Communications (CFIC), 2013 Conference on}.
  IEEE, 2013.

\bibitem{wischik2011design}
D.~Wischik, C.~Raiciu, A.~Greenhalgh, and M.~Handley.
\newblock Design, implementation and evaluation of congestion control for
  multipath tcp.
\newblock In {\em Usenix NSDI}, 2011.

\bibitem{polyzos}
G.~Xylomenos, C.~Ververidis, V.~Siris, N.~Fotiou, C.~Tsilopoulos, X.~Vasilakos,
  K.~Katsaros, and G.~Polyzos.
\newblock A survey of information-centric networking research.
\newblock {\em Communications Surveys Tutorials, IEEE}, PP(99):1--26, 2013.

\bibitem{yi2013case}
C.~Yi, A.~Afanasyev, I.~Moiseenko, L.~Wang, B.~Zhang, and L.~Zhang.
\newblock A case for stateful forwarding plane.
\newblock {\em Computer Communications}, 36(7):779--791, 2013.

\end{thebibliography}

\section*{Appendix I: CCNx implementation details}

This appendix reports some implementation details. The source code is available at \cite{software} and it is based on CCNx v0.7.2; our main contributions are in ccnd.c, for the multipath forwarding strategies used by every node, and in CCNAbstractInputStream.java for the AIMD congestion control used by the receiver.

\emph{Path identification} -  The multipath forwarding strategy of a node identifies the network paths used by content transfers in progress only through the output faces. Indeed, there is no explicit indication of path diversity in the current implementation of CCNx, besides the face number. Thus, if for a given content multiple paths are available beyond the same face, a strategy would consider all such paths as a single path. 

\emph{Path monitoring} - The strategy monitors the status of the paths by using a dynamic-table which, for each content transfer in progress, and for each face, stores the number of pending Interest messages and a smoothed average of RTT samples. For a given content, the strategy derives RTT samples by considering some forwarding Interest messages as RTT probes. When no RTT probe is pending on a given face, the next Interest sent on that face takes the role of a RTT probe. When the Data message related to the RTT probe comes back, a sample of the path RTT is evaluated as the difference between the receiving time of the Data message and the sending time of the RTT probe.

\emph{Loss detection} -  The AIMD congestion control detects the loss of a Data message by means of an Interest timeout. The timeout computation is the one provided by the CCNx implementation, which evaluates the timeout as a smoothed average of RTT samples observed by the congestion control process, multiplied by a RTT\_FACTOR (e.g. 2). The RTT\_FACTOR takes into account the delay variations among different network paths. We piggybacked the timeout value in the LIFETIME field of Interest messages. The multipath forwarding strategy detects the loss of a Data message on a path by means of an Interest timeout equal to the LIFETIME field. 

\emph{Evaluation of the pipeline capacity for the FPF strategy} - Each node estimates the pipeline capacity of the path beyond a face as the number of Interest messages in-flight when a Data loss is detected, multiplied by a guard factor equal to 3/4. Since a node does not discern the presence of multiple paths beyond the next hop node, the pipeline capacity measured by a node can be initially lower than the real one; however, after some loss rounds, this underestimation error decreases.

\end{document}